   \pdfoutput=1
   \documentclass[fleqn,usenatbib,usedcolumn]{mnras}
   \usepackage[british]{babel}             
   \let\la=\lesssim 
   \usepackage[T1]{fontenc}                
   \usepackage{graphicx}                   
   \hypersetup{pdfauthor={J. G. Hj{\o}rringgaard},
               pdftitle={Testing stellar evolution models with the retired A star HD 185351},
               pdfkeywords={asteroseismology -- stars: individual (HD 185351) -- stars: interiors},
               bookmarksnumbered=true}
   \volume{{\rm in press}}
   
\usepackage{amsmath}	
\usepackage{amssymb}	
\usepackage{siunitx}	
\usepackage{times,txfonts}

\usepackage{xcolor}
\defcitealias{Johnson2014}{J14}

\newcommand{\FEH}{\left[\mathrm{Fe/H}\right]}

\usepackage{pgfplots}
\newlength\figureheight 
\newlength\figurewidth 

\title[The retired A star HD 185351]{Testing stellar evolution models with the retired A star HD 185351}
\author[J. G. Hj{\o}rringgaard et al.]{J. G. Hj{\o}rringgaard,$^{1}$\thanks{E-mail: jgh@phys.au.dk}
V. Silva Aguirre,$^{1}$
T. R. White,$^{1}$
D. Huber,$^{2,1,3}$
B. J. S. Pope,$^{4}$ \newauthor
L. Casagrande,$^{5}$
A. B. Justesen,$^{1}$
and J. Christensen-Dalsgaard$^{1}$
\\
$^{1}$Stellar Astrophysics Centre, Department of Physics and Astronomy, Aarhus University, Ny Munkegade 120, DK-8000 Aarhus C, Denmark\\
$^{2}$Sydney Institute for Astronomy (SIfA), School of Physics, University of Sydney, NSW 2006, Australia\\
$^{3}$SETI Institute, 189 Bernardo Avenue, Mountain View, CA 94043, USA\\
$^{4}$Oxford Astrophysics, University of Oxford, Keble Rd, Oxford OX1 3RH, UK\\
$^{5}$Research School of Astronomy and Astrophysics, Mount Stromlo Observatory, The Australian National University, ACT 2611, Australia
}

\date{Accepted XXX. Received YYY; in original form ZZZ}

\pubyear{2016}

\begin{document}
\label{firstpage}
\pagerange{\pageref{firstpage}--\pageref{lastpage}}
\maketitle

\begin{abstract}
The physical parameters of the retired A star HD 185351 were analysed in great detail by \citet{Johnson2014} using interferometry, spectroscopy and asteroseismology. Results from all independent methods are consistent with HD 185351 having a mass in excess of $1.5\mathrm{M}_{\sun}$. However, the study also showed that not all observational constraints could be reconciled in stellar evolutionary models, leading to mass estimates ranging from $\sim 1.6-1.9\mathrm{M}_{\sun}$ and casting doubts on the accuracy of stellar properties determined from asteroseismology. Here we solve this discrepancy and construct a theoretical model in agreement with all observational constraints on the physical parameters of HD 185351. The effects of varying input physics are examined as well as considering the additional constraint of the observed g-mode period spacing. This quantity is found to be sensitive to the inclusion of additional mixing from the convective core during the main sequence, and can be used to calibrate the overshooting efficiency using low-luminosity red giant stars. A theoretical model with metallicity $\left[\mathrm{Fe/H}\right]=0.16$dex, mixing-length parameter $\alpha_{\mathrm{MLT}}=2.00$, and convective overshooting efficiency parameter $f=0.030$ is found to be in complete agreement with all observational constraints for a stellar mass of $M\simeq1.60\mathrm{M}_{\sun}$.
\end{abstract}



\section{Introduction}\label{sec:Introduction}
The red-giant star HD185351 is an important benchmark star for testing model-dependent mass estimates for evolved stars.
Its physical properties were thoroughly examined by \citet[][hereafter J14]{Johnson2014}, with an analysis of properties obtained from interferometry, spectroscopy, and asteroseismology. From spectroscopy the metallicity was determined to be $\left[\mathrm{Fe/H}\right]=0.16\pm 0.04$dex and the effective temperature $T_{\mathrm{eff}}(\mathrm{SME})=5016\pm 44\si{\kelvin}$, while interferometry yielded $T_{\mathrm{eff}}(\mathrm{INTF})= 5042\pm 32\si{\kelvin}$ (or $T_{\mathrm{eff}}=5047\pm 48\si{\kelvin}$ when using a $F_{\mathrm{bol}}$ determination from the \citet{Casagrande2010} implementation of the infrared flux method), and a radius of $R=4.97\pm 0.09\mathrm{R}_{\sun}$. The asteroseismic parameters obtained were the frequency of maximum power $\nu_{\mathrm{max}}=229.8 \pm 6.0\si{\micro\hertz}$, large frequency separation $\Delta\nu = 15.4\pm 0.2\si{\micro\hertz}$, and the g-mode period spacing $\Delta\Pi_1 = 104.7\pm 0.2\si{\second}$ (a thorough review of the observational methods can be found in \citetalias{Johnson2014}). By interpolating the spectroscopic parameters onto Yonsei-Yale model grids \citetalias{Johnson2014} obtained a mass of $1.87\pm0.07\mathrm{M}_{\sun}$, which was found to be in good agreement with the asteroseismic mass estimated from $\nu_{\mathrm{max}}$ and $\Delta\nu$, ($1.99\pm 0.23\mathrm{M}_{\sun}$), but in tension to the most model-independent estimate based on the combination of $\Delta\nu$ with the interferometric radius ($1.60\pm 0.08\mathrm{M}_{\sun}$). \citetalias{Johnson2014} concluded that this discrepancy could either be explained by small systematic errors in the observations, or systematic errors in the input physics of the adopted models.

The results of \citetalias{Johnson2014} are consistent with a mass in excess of $1.5\mathrm{M}_{\sun}$, indicating that HD 185351 was an early F- or A-type star on the main sequence (MS). However mass estimates of single stars are based on stellar evolution models which may contain systematic errors, and indeed the accuracy of mass estimates of sub-giant stars has recently been called into question \citep{Lloyd2011,Lloyd2013,Schlaufman2013}, suggesting that sub-giants with masses in excess of $1.5\mathrm{M}_{\sun}$ should be rare in the solar neighbourhood. This issue of mass estimation for this kind of star has implications for the reliability of stellar evolution models as well as our understanding of planet occurrence rates around higher-mass stars which depends on accurate knowledge of the stellar masses \citep{Lloyd2013,Johnson2013}. Recent efforts to resolve the mass problem by comparing spectroscopy to independent results (from binary systems or asteroseismology) suggests that masses of evolved stars determined from evolutionary grids are not significantly affected by systematic errors \citep{Ghezzi2015b}.

The scope of the paper is to test if a stellar model can be constructed in order to agree with all of the observational constraints, and explore whether the model-dependent input physics can have a significant effect on the inferred masses. We begin in Section~\ref{sec:LongCadence} by determining an updated value of $\nu_{\mathrm{max}}$ from \textit{Kepler} long-cadence data. Section~\ref{sec:Method} outlines the methods used in the present work to compute stellar evolution tracks, and extract asteroseismic parameters from these, to compare with the observations. A possible resolution of the problem at hand by corrections to the asteroseismic scaling relations is presented and discussed. The computed models with varying input physics, including changes in metallicity, mixing-length, and inclusion of convective overshooting, are presented in Section~\ref{sec:Results}. The results are discussed in Section~\ref{sec:Discussion}, where a combination of the individual variations in the model input is used in order to construct a model, with reasonable input physics, which is consistent with all observational constraints within a region of the stellar evolution grid. 
\section{Asteroseismic data analysis}\label{sec:LongCadence}
The measurement of asteroseismic parameters by \citetalias{Johnson2014} was made from \textit{Kepler} short-cadence (SC; one-minute sampling) photometry. A value for the frequency of maximum power, $\nu_{\mathrm{max,SC}}=229\pm 6.0 \si{\micro\hertz}$ was obtained.
These observations only span a single \textit{Kepler} observing Quarter (Q16), with a duration of 85.6 days. With the stochastic nature of the excitation and damping of solar-like oscillations, mode heights vary over short timescales, and this has the potential to affect the measurement of $\nu_{\mathrm{max}}$. Ideally, longer observations would be desirable, but with the failure of two reaction wheels causing the end of the nominal \textit{Kepler} mission, further observations of HD 185351 are no longer possible.

\begin{figure}
\centering
\includegraphics[width=1\columnwidth]{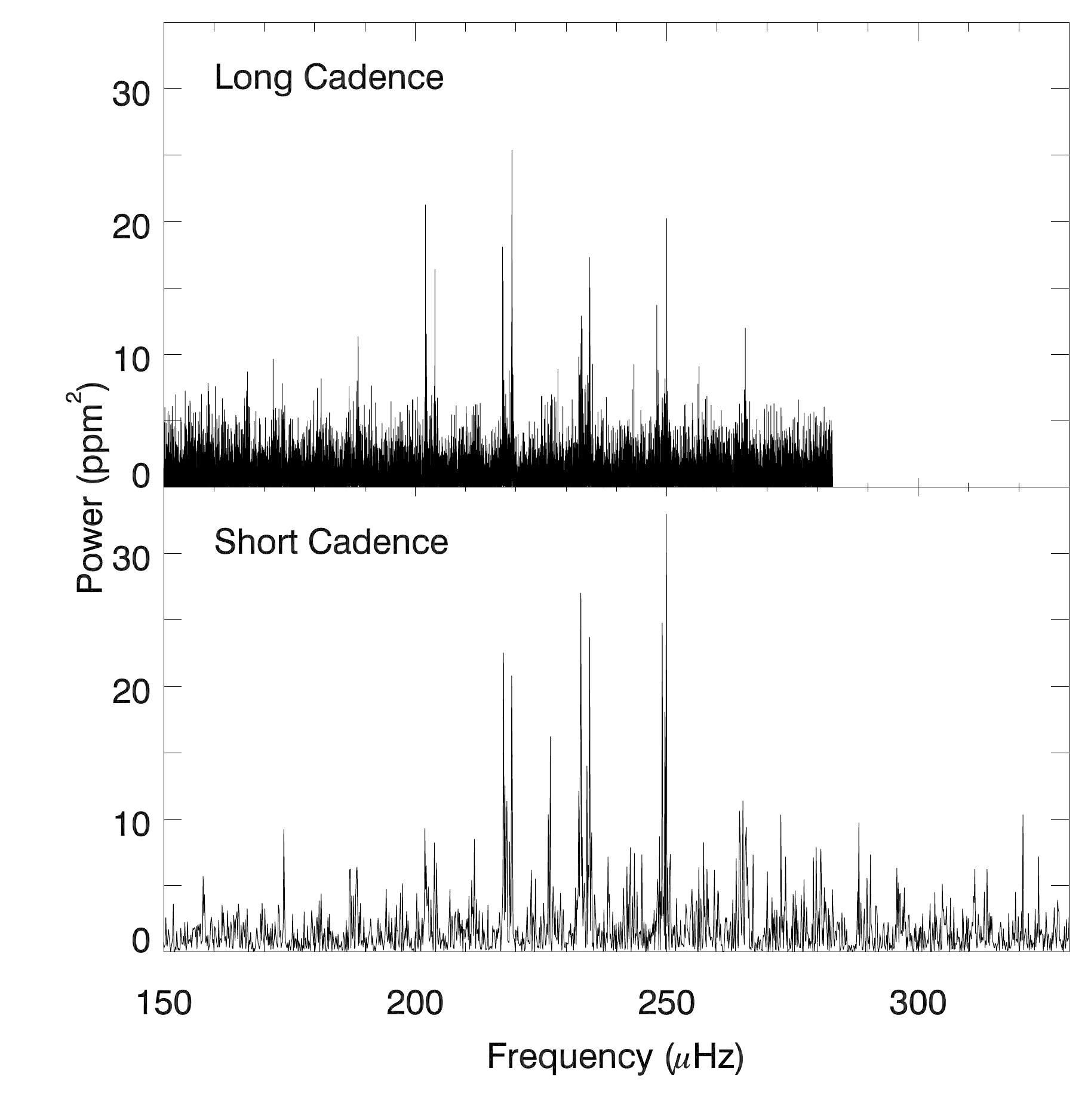}
\caption{\textit{Top panel}: Background-corrected power spectrum of HD 185351 obtained from the smear calibrated \textit{Kepler} LC light curve. \textit{Bottom panel}: Background-corrected power spectrum of HD 185351 obtained from \textit{Kepler} SC light curve of Q16.}
\label{fig:LC}
\end{figure}

Fortunately, \citet{Pope2016} have recently shown how long-cadence (LC; 30-minute sampling) light curves may be recovered for bright stars from \textit{Kepler} `smear' calibration data. We have applied this method to HD 185351 to obtain the full 4-year LC light curve, and the resulting power spectrum is shown in Fig.~\ref{fig:LC} where it is compared to the SC power spectrum.

Although the smear light curve has a greater frequency resolution than the original SC data, the noise level is higher, and the modes have attenuated amplitudes because they are just below the LC Nyquist frequency of $\sim 283\si{\micro\hertz}$. The LC smear light curve therefore provides limited opportunity to improve the asteroseismic analysis of \citetalias{Johnson2014}. Therefore we use the values of $\Delta\nu$ and $\Delta\Pi_1$ determined by \citetalias{Johnson2014}.  For $\nu_{\mathrm{max}}$, however, we have made revised measurement with the smear data using the method by \citet{Huber2009}. To account for Nyquist attenuation, the power spectrum was corrected using \citep{Murphy2012}
\begin{equation}
A_{\mathrm{cor}} = A\cdot \mathrm{sinc} \left(\dfrac{\pi}{2}\cdot \dfrac{\nu}{\nu_{\mathrm{Nyq}}}\right)
\end{equation}
where $A$ and $A_{\mathrm{cor}}$ are the original and corrected amplitude, respectively, and $\nu_{\mathrm{Nyq}}$ is the long-cadence Nyquist frequency. The analysis yielded $\nu_{\mathrm{max}}=218.9\pm 4.8\si{\micro\hertz}$, which is $1.4\sigma$ smaller than the value determined from the SC data using the same method. Owing to the difficulties of measuring $\nu_{\mathrm{max}}$ close to the Nyquist frequency, balanced against the benefit of using the longer time series, we adopt a range for $\nu_{\mathrm{max}}$ that encompasses both the SC and LC values, $\left(214.1,\, 235.8\right) \si{\micro\hertz}$.
\section{Method}\label{sec:Method}
The computation of stellar evolution tracks is done using the Garching Stellar Evolution Code \citep[\texttt{GARSTEC},][]{GARSTEC}. All tracks are computed with the 2005 OPAL equation of state \citep{Rogers1996,Rogers2002} extended with MHD equation of state \citep{Mihalas1988} in low temperature regions. The opacity tables used are the OPAL opacities for high temperatures \citep{Iglesias1996} and those of \citet{Ferguson2005} for low temperatures. Energy transport by convection is treated by the mixing-length theory as described by \citet{Kippenhahn2012}.

The physical and asteroseismic parameters can be extracted from these tracks in order to obtain regions where the model values agree with the observed ones. The asteroseismic parameters are not standard output of \texttt{GARSTEC} and thus were computed independently.
\subsection{Asteroseismology}
The frequency of maximum power $\nu_{\mathrm{max}}$ has been proposed to scale with the acoustic cutoff frequency $\nu_{\mathrm{c}}$ \citep{Brown1991}, such that $\nu_{\mathrm{max}}\propto \nu_{\mathrm{c}} \propto c/H_p$, where $c$ and $H_P$ are the adiabatic sound speed and pressure scale height in the star's photosphere. Assuming an ideal gas the scaling relation for the frequency of maximum power can be expressed as \citep[e.g.][]{Kjeldsen1995}
\begin{equation}
\nu_{\text{max}} = \nu_{\text{max},\sun}\left(\dfrac{M}{\mathrm{M}_{\sun}}\right)\left(\dfrac{R}{\mathrm{R}_{\sun}}\right)^{-2}\left(\dfrac{T_{\text{eff}}}{\mathrm{T}_{\mathrm{eff},\sun}}\right)^{-1/2}. \label{eq:numax_sca}
\end{equation}

Through asymptotic analysis of the equations of adiabatic oscillations \citep[e.g.][]{Deubner1984} the large frequency separation can be expressed as 
\begin{equation}
\Delta\nu = \left(2\int_{0}^{R}\dfrac{\mathrm{d}r}{c}\right)^{-1}. \label{eq:ss_int}
\end{equation}
From this expression it can be obtained by homology relations that the large frequency separation is related to the mean density $\bar{\rho}$ of the star by $\Delta\nu\propto\bar{\rho}^{1/2}$ \citep{Ulrich1986}. Scaling with respect to the solar p-mode spectrum yields 
\begin{equation}
\Delta\nu = \Delta\nu_{\sun}\left(\dfrac{M}{\mathrm{M}_{\sun}}\right)^{1/2}\left(\dfrac{R}{\mathrm{R}_{\sun}}\right)^{-3/2}.\label{eq:Dnu_sca}
\end{equation}
For the analysis the adopted solar values are $\Delta\nu_{\sun}=135.1\si{\micro\hertz}$ and $\nu_{\mathrm{max},\sun}=3090\si{\micro\hertz}$ \citep{Huber2011}. 

A third way to compute the large frequency separation is by modelling the adiabatic oscillations. In this work this is done using the Aarhus Adiabatic Oscillation Package \citep[\texttt{ADIPLS},][]{ADIPLS}, which calculates the oscillation frequencies directly from the evolutionary models. The large frequency separation is calculated from the oscillation frequencies by a least-squares fit to the frequencies as a function of radial order $n$, weighting the frequencies with a Gaussian envelope centred at the frequency of maximum power obtained from the scaling relation in Equation~\eqref{eq:numax_sca} with a full width at half maximum of $0.25\nu_{\mathrm{max}}$ as proposed by \citet{White2011}.

The asymptotic treatment shows that p-modes are equally spaced in frequency by the large frequency separation given in Equation~\eqref{eq:ss_int}. By a similar analysis \citep[e.g.][]{Tassoul1980} the g-modes have been shown to be equally spaced in period, with the period spacing 
\begin{equation}
\Delta\Pi_1 = \dfrac{\Pi_0}{L}, \label{eq:DPi}
\end{equation}
where $L^2 = l(l+1)$, and
\begin{equation}
\Pi_0 = 2\pi^2\left(\int_{r_1}^{r_2}N\dfrac{\mathrm{d}r}{r}\right)^{-1}. 
\end{equation}
Here $N$ is the Brunt-V{\"a}is{\"a}l{\"a} frequency given by
\begin{equation}
 N^2 = g\left(\dfrac{1}{\Gamma_1}\dfrac{\mathrm{d}\ln p}{\mathrm{d}r} - \dfrac{\mathrm{d}\ln\rho}{\mathrm{d}r}\right), \label{eq:N}
\end{equation} 
where $g$ is the gravitational acceleration, $p$ is the pressure, $\rho$ is the density, and the adiabatic exponent is given by $\Gamma_1 = (\partial\ln p/\partial\ln\rho )_{\mathrm{ad}}$. 

We calculated $\Delta\nu$ using both the scaling relation (eq. 2) and from individual frequencies computed using \texttt{ADIPLS}. We found good agreement between the values of $\Delta\nu$ computed using the two methods. We also applied a metallicity-dependent correction to the scaling relation (see Serenelli, in prep). However, due to the location of HD 18531 in the Hertzsprung-Russell diagram (HRD), the correction is negligible in this case. For the remainder of this paper, we chose to use $\Delta\nu$ from the corrected scaling relation. The conclusions of the paper are not dependent on this choice.

\subsection{Kiel diagram}
\begin{figure}
	\includegraphics[width=\columnwidth]{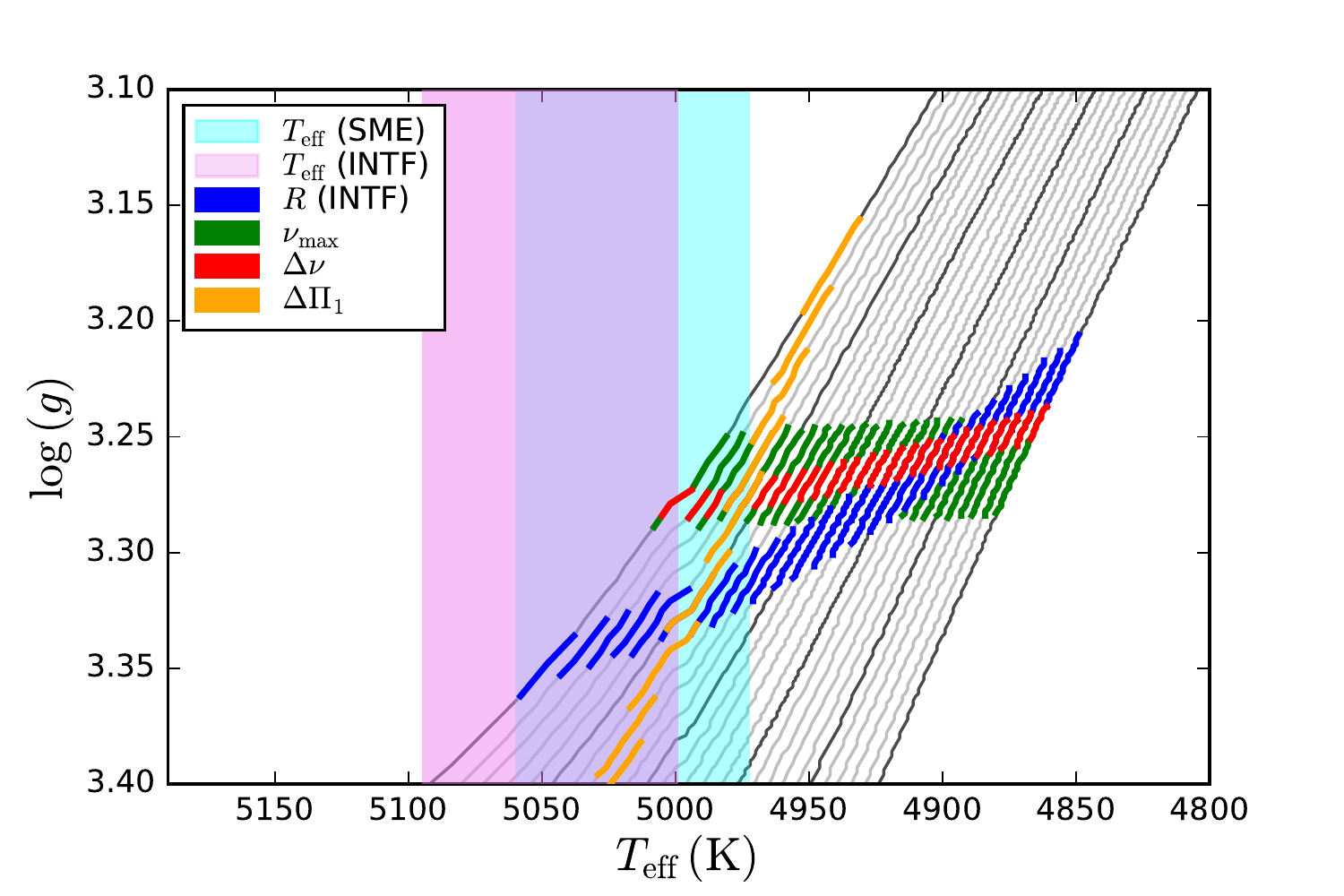}
    \caption{Surface gravity plotted against effective temperature for a set of evolutionary tracks, with $\left[\mathrm{Fe/H}\right]=0.16$ dex and no convective overshooting, computed using \texttt{GARSTEC}. The coloured bands show $1\sigma$ ($10\sigma$ for $\Delta\Pi_1$) constraints obtained from observations primarily by \citetalias{Johnson2014} (see Section~\ref{sec:Introduction} for values). Black tracks correspond to masses $1.5-2.0\mathrm{M}_{\sun}$ (from right to left) in steps of $0.1\mathrm{M}_{\sun}$. See text for details.}
    \label{fig:standard}
\end{figure}

Figure~\ref{fig:standard} shows the computed $\log g$-$T_{\mathrm{eff}}$ diagram (commonly called Kiel diagram) with metallicity $\left[\mathrm{Fe}/\mathrm{H}\right]=0.16$ as obtained from spectroscopy\footnote{The small break in the evolutionary tracks at $T_{\mathrm{eff}}=5000\si{\kelvin}$ is due to transitions between equation of state tables.}. The different colours in Fig.~\ref{fig:standard} represent the various $1\sigma$ constraints from observations (the constraint band for $\Delta\Pi_1$ is widened to $10\sigma$ in order to make it visible in the diagram -- a $1\sigma$ band would naturally lie within this band, and the conclusions are not affected by the additional width; see Section~\ref{sec:Introduction} for values).

Compared to the corresponding Kiel diagram by \citetalias{Johnson2014}, where BaSTI evolutionary tracks are used, there is an additional constraint on the models applied to the grid in Fig.~\ref{fig:standard} -- the g-mode period spacing $\Delta\Pi_1$ displayed as a orange band. The location of this additional asteroseismic constraint is not in agreement with both the other asteroseismic constraints and the interferometric radius at one region of the Kiel diagram, indicating that the theoretical description of the star is incomplete. $\Delta\nu$ and $\nu_{\mathrm{max}}$ are consistent with the interferometric radius at a stellar mass of $M\simeq 1.6\mathrm{M}_{\sun}$, although for models significantly cooler than what is observed. These discrepancies between the locations of the different constraint bands in the stellar evolution grid of Fig.~\ref{fig:standard} motivate further study of the influence of the model input on the grid in general and the constraint bands within.

In Section~\ref{sec:LongCadence} we discussed the individual reliability of $\nu_{\mathrm{max}}$ determined from \textit{Kepler} SC and LC photometry. It is not possible wih the current data to determine which value is most appropriate to use, and thus the constraint applied to the evolutionary tracks is a combination of the two in the range $\left(214.1, 235.8\right)\si{\micro\hertz}$ (green band in Fig.~\ref{fig:standard}). As is evident from the Kiel diagram, this constraint on $\nu_{\mathrm{max}}$ is of little use for determining the appropriate model to reproduce the stellar parameters of HD 185351. The $\nu_{\mathrm{max}}$ constraint agrees with the $\Delta\nu$ and $\Delta\Pi_1$ constraints for a model mass of $M\simeq 1.92\mathrm{M}_{\sun}$, while it agrees with the $\Delta\nu$ and $R$ constraints at a model mass of $M\simeq 1.62\mathrm{M}_{\sun}$. No additional information is added by including the constraint on $\nu_{\mathrm{max}}$ in the grid, therefore it will not be depicted in the remaining Kiel diagrams in this paper.
\section{Results}\label{sec:Results}
Variations in different sets of input physics of the computed stellar evolution tracks influence the location of the tracks, and the constraint bands, in different ways. One of the basic properties of a star needed by stellar evolution models to produce tracks is the metallicity, defined as $\left[\mathrm{Fe/H}\right]=\log\left(Z/X\right)-\log\left(Z/X\right)_{\sun}$ (where $X$ and $Z$ are the mass fractions of hydrogen and elements heavier than helium respectively), which is found to be $\left[\mathrm{Fe/H}\right]=0.16\pm 0.04$ dex by \citetalias{Johnson2014}. Within the range of the uncertainty of the metallicity determination there is some variation in computed evolutionary tracks. Other variations in the input physics of the models that we consider here are the inclusion of overshooting of mass elements from convective regions and variations in the mixing-length parameter $\alpha_{\mathrm{MLT}}$. A summary of the various computed models are listed in Table~\ref{tab:figures}, for quick reference to the figures. 
\begin{table}
\centering
\caption{Summary of input physics of computed models. Here $f$ is the overshooting efficiency parameter and $\alpha_{\mathrm{MLT}}$ is the mixing-length parameter. All grids are in the mass range of $1.5-2.0\mathrm{M}_{\sun}$ with mass separation $0.02\mathrm{M}_{\sun}$.}
\begin{tabular}{ccccc}
\hline
Figure  & $\left[\mathrm{Fe/H}\right]$ & $f$ & $\alpha_{\mathrm{MLT}}$ \\
\hline
\ref{fig:standard} & $0.16$ & $0$ & $1.791$ \\
\ref{fig:f0016} & $0.16$ & $0.016$ & $1.791$ \\
\ref{fig:FeH} & $0.08$ & $0$ & $1.791$ \\
\ref{fig:alpha2.0} & $0.16$ & $0$ & $2.00$ \\
\ref{fig:BestFit_alpha} & $0.16$ & $0.030$ & $2.00$ \\
\hline
\end{tabular} 
\label{tab:figures}
\end{table}
\subsection{Convective overshooting}\label{sec:Overshooting}
There is plenty of evidence in the literature that convective core sizes predicted from models using the Schwarzschild criterion alone are underestimating the actual mixed core extent \citep[e.g.][]{Maeder1974,Maeder1991,Chiosi1992,Vandenberg2007}, and recent analyses of \textit{Kepler} stars support this observation \citep{SilvaAguirre2013,Deheuvels2016}. Therefore it is reasonable to assume that some amount of mixing beyond the formal Schwarzschild convective boundary should be included in the calculation of the evolutionary tracks. In particular, the inclusion of convective overshooting increases the mixing of elements in the core, extending the lifetime of the star on the MS. \texttt{GARSTEC} treats convective overshooting as a diffusive process, as described by \citet{Freytag1996,Herwig1997}, with diffusion constant
\begin{equation}
D(z) = D_0 \exp \left(\dfrac{-2z}{fH_p}\right), \label{eq:overshooting}
\end{equation}
where $D_0$ sets the scale of diffusive speed and is derived from the convective velocity in the mixing-length theory, $z$ is the radial distance to the Schwarzschild border, $H_p$ is the pressure scale height, and $f$ is the overshooting efficiency parameter. In \texttt{GARSTEC} the value for $f$ has been calibrated to $0.016$ by isochrone fitting to the color-magnitude diagram of M67 \citep[see,][]{Magic2010}.
\begin{figure}
	\includegraphics[width=\columnwidth]{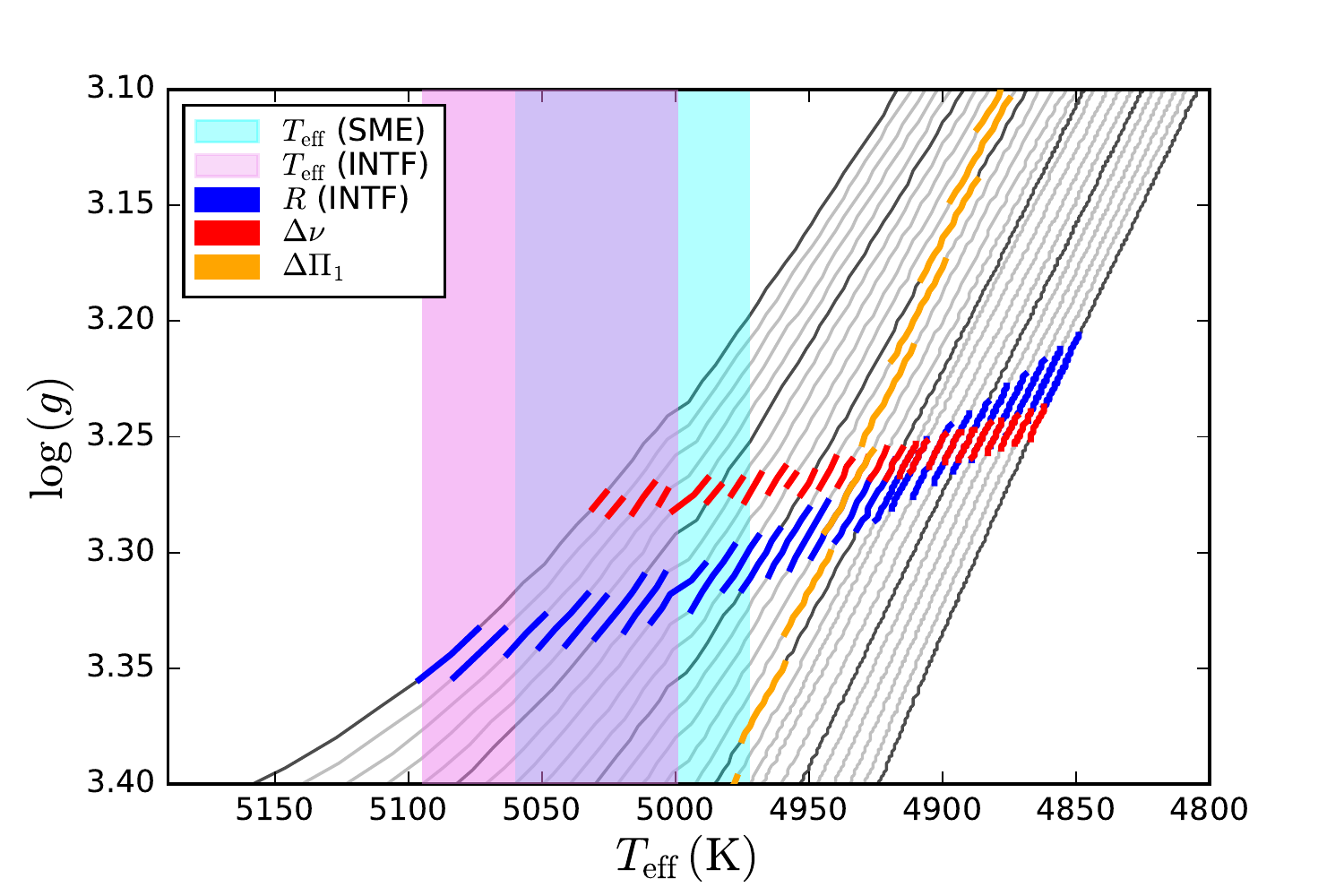}
    \caption{Kiel diagram computed for a model with $\left[\mathrm{Fe/H}\right]=0.16$ dex and convective overshooting included at $f=0.016$. The colour code is as in Fig.~\ref{fig:standard}.}
    \label{fig:f0016}
\end{figure}

The computed Kiel diagram with metallicity $\left[\mathrm{Fe/H}\right] = 0.16$ dex and overshooting efficiency parameter $f=0.016$ is shown in Fig.~\ref{fig:f0016}. The locations of the constraint bands are similar except for the g-mode period spacing, which is shifted significantly towards cooler regions, where the $\Delta\nu$ and radius constraints agree. Thus matching $\Delta\Pi_1$ with these other parameters can serve as a test for convective overshoot during the MS. The deviation from the observed temperature is still an issue for this set of input physics, and will be resolved by other input variations.

The added constraint of the g-mode period spacing $\Delta\Pi_1$ is extremely sensitive to the overshooting efficiency parameter of the model, and the exact cause for this is a change in the thermodynamic state in the centre of the star. The direct impact of including overshooting is increased mixing in the core, varying the relationship between density and pressure gradients, making the Brunt-V{\"a}is{\"a}l{\"a} frequency change (see Equation~\eqref{eq:N}).
\begin{figure}
\centering
\includegraphics[width=\columnwidth]{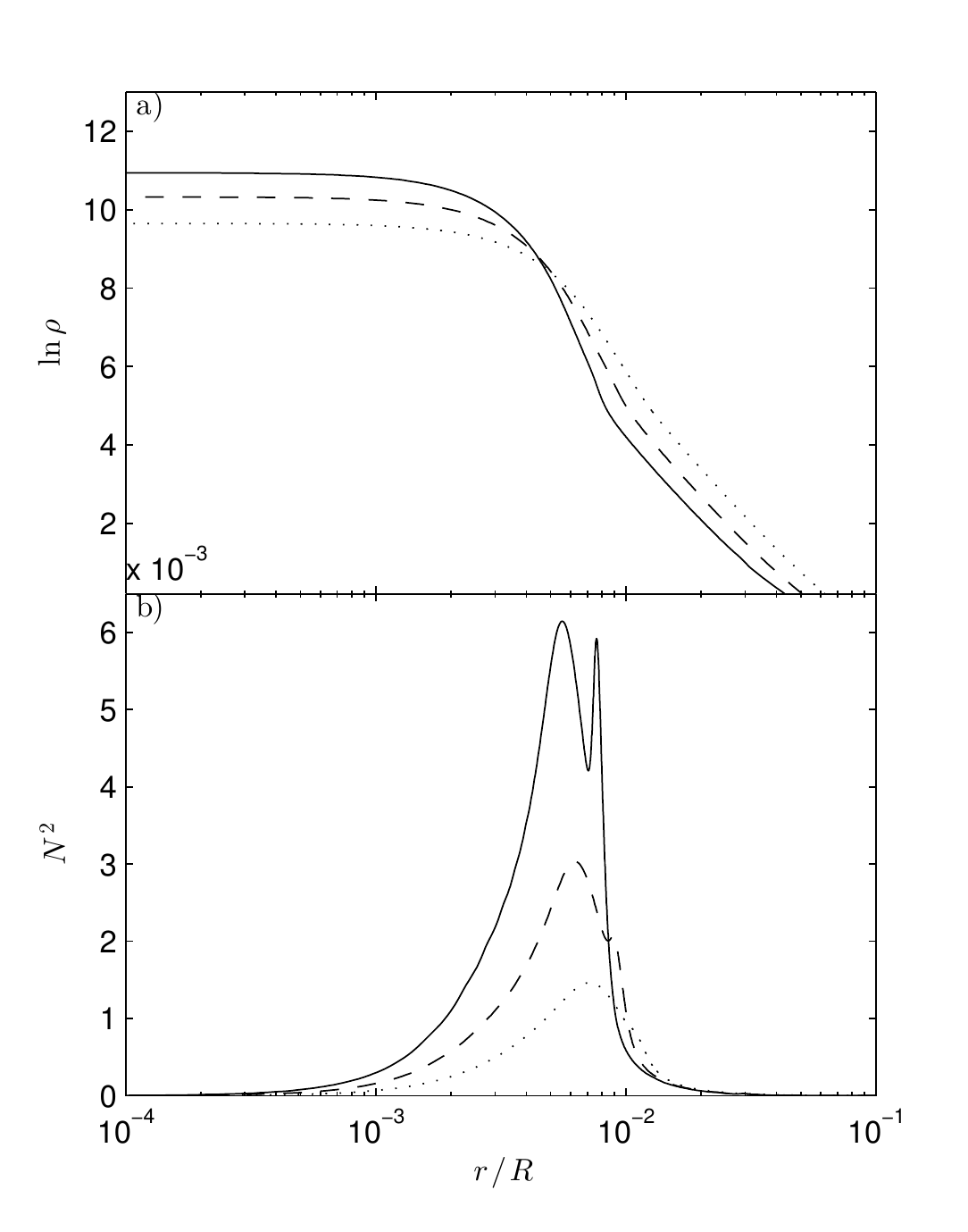}
\caption{\textit{Top panel}: Density $\rho$ as a function of the radial distance to the core for three $1.90\mathrm{M}_{\sun}$ models selected at the observed $\Delta\nu$ value with metallicity $\left[\mathrm{Fe/H}\right]=0.16$ dex, and $f=0$ (solid), $f=0.016$ (dashed), and $f=0.032$ (dotted). \textit{Bottom panel}: Brunt-V{\"a}is{\"a}l{\"a} frequency profile for the same models.}
\label{fig:RHOandN}
\end{figure}

Figure~\ref{fig:RHOandN} shows the density (top panel) and Brunt-V{\"a}is{\"a}l{\"a} frequency profile (bottom panel) for three models of varying amount of overshooting. The large impact on the Brunt-V{\"a}is{\"a}l{\"a} frequency arises from a steeper density gradient when overshooting is not included. A star of this mass has a convective core which decreases in size during the MS, and inclusion of overshooting in this phase alters the density profile of the star at later stages of the evolution.
\subsection{Metallicity}
\begin{figure}
	\includegraphics[width=\columnwidth]{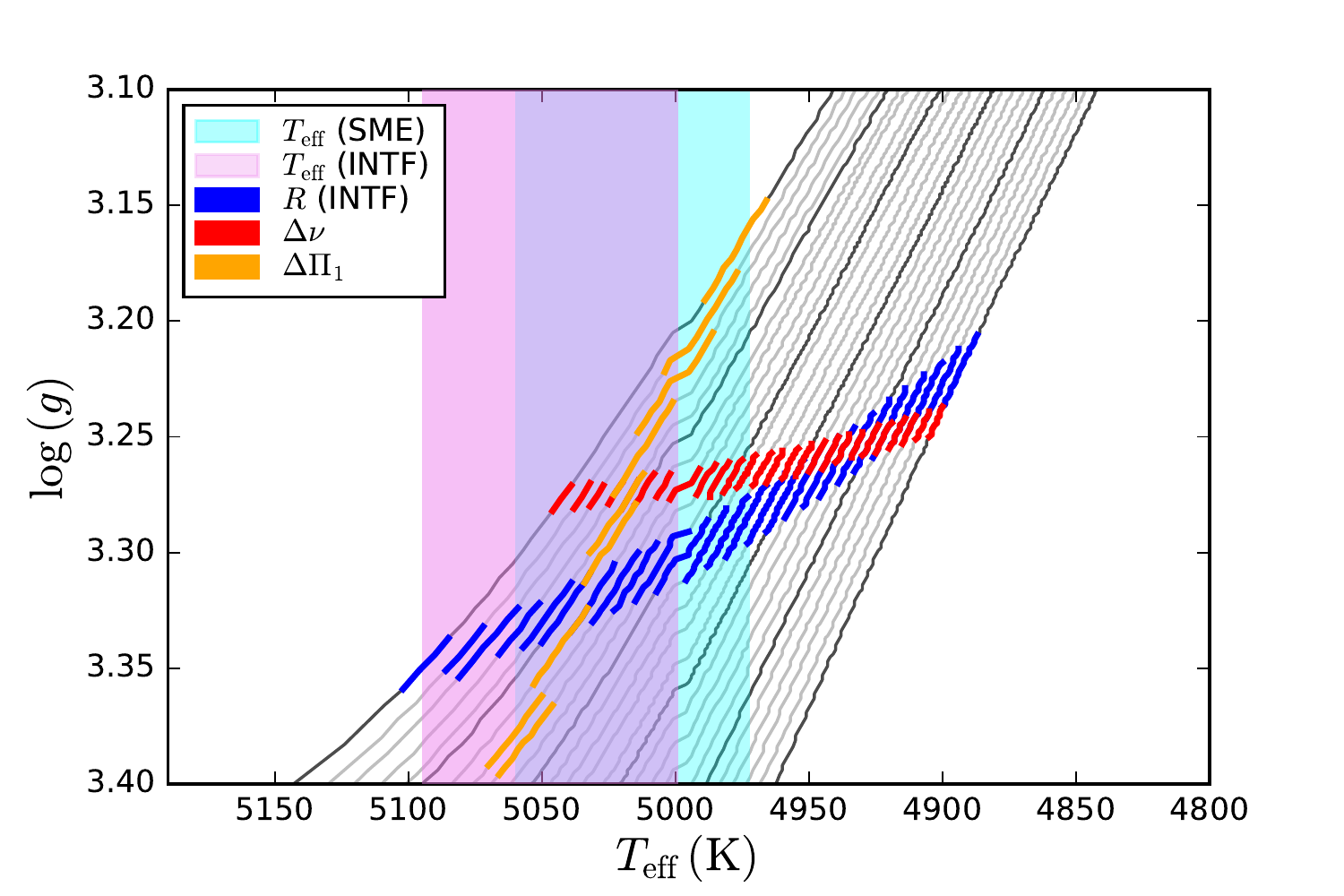}
    \caption{Kiel diagram computed with metallicity $\left[\mathrm{Fe/H}\right]=0.08$ dex and no inclusion of convective overshooting. The colour code is as in Fig.~\ref{fig:standard}.}
    \label{fig:FeH}
\end{figure}

The primary effect on the computed stellar evolution grid from varying the metallicity is a shift toward hotter regions for a decrease in $\left[\mathrm{Fe/H}\right]$. This effect is visible when comparing the Kiel diagrams in Fig.~\ref{fig:standard} and \ref{fig:FeH}, which are computed with metallicities of $\left[\mathrm{Fe/H}\right]=0.16$ and $0.08$ dex respectively. The $2\sigma$ variation in metallicity is chosen as a conservative approach \citep[recent observations support this, with a metallicity of $\FEH=0.10\pm 0.04$ reported by][]{Ghezzi2015}.
The locations of the constraint bands from observations in the two Kiel diagrams coincide (see Fig.~\ref{fig:standard} \& \ref{fig:FeH}), in the sense that asteroseismic parameters and the interferometric radius constraints are shifted with the evolutionary tracks towards hotter regions,
making the zone where $\Delta\nu$ and radius constraints agree approach the observed temperatures. The temperature differences are, however, not completely resolved by the applied decrease in metallicity, just as the discrepancy between the locations of the individual asteroseismic constraint bands remains. For an increase in metallicity the effect on the computed stellar evolution grid is a shift toward cooler temperatures. Thus the parameters are further from being reconciled.
\subsection{Mixing-length parameter}
From observations of globular clusters and field stars it is known that there is a discrepancy between temperatures observed in the red-giant branch and those predicted from stellar evolution models \citep[see e.g.,][]{Ferraro1999,Vandenberg2008,Salaris2008,Pinsonneault2014}. The origin of this discrepancy could lie on the incomplete treatment of convection in hydrostatic stellar evolution models, usually described with the mixing-length formalism. In this framework convective motions are described with the mixing-length parameter $\alpha_{\mathrm{MLT}}$ normally calibrated to reproduce the solar properties.
\begin{figure}
	\includegraphics[width=\columnwidth]{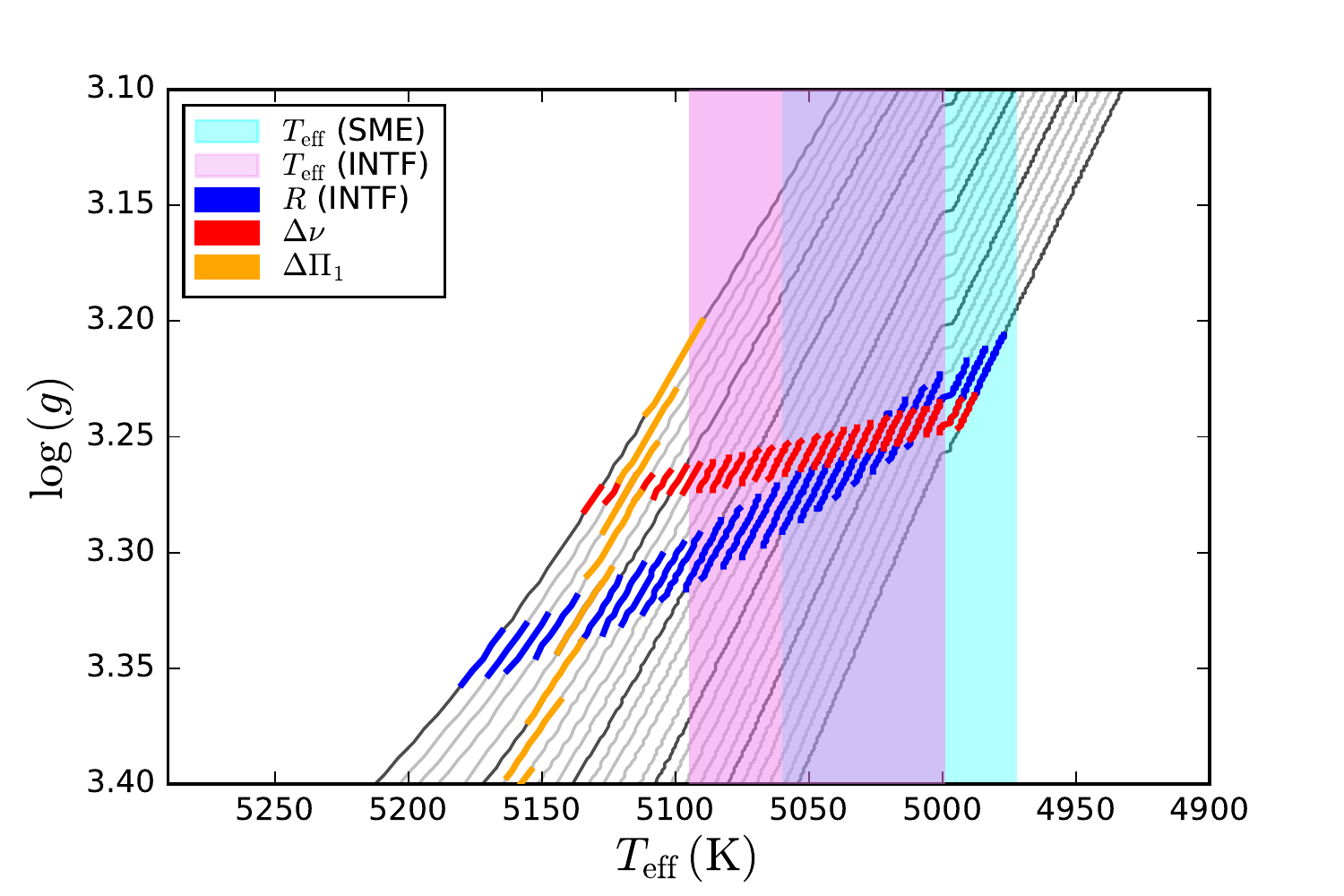}	
    \caption{Kiel diagram computed with $\left[\mathrm{Fe/H}\right]=0.16$ dex, no convective overshooting, and a mixing-length parameter of $\alpha_{\mathrm{MLT}} = 2.00$. The colour code is as in Fig.~\ref{fig:standard}.}
    \label{fig:alpha2.0}
\end{figure}
The value of $\alpha_{\mathrm{MLT}}$ does however vary throughout the HRD \citep{Trampedach2014,Magic2015}, and the discrepancy between the different constraints in the Kiel diagram could originate from an alternate value of $\alpha_{\mathrm{MLT}}$ being more suitable for the star.

The question of the universality of the mixing-length parameter has also been investigated using the binary system $\alpha$ Centauri showing that different values of $\alpha_{\mathrm{MLT}}$ are required to match observations with standard stellar models \citep{Demarque1986,Miglio2005,Yildiz2007}. Tests using eclipsing binaries \citep{Ludwig1999}, clusters \citep{Yildiz2006}, and \textit{Kepler} data \citep{Bonaca2012} also suggest that the mixing-length parameter should vary with stellar properties. 

In this paper a value of $\alpha_{\mathrm{MLT}}=1.791$ has been used as the standard solar-calibrated value in \texttt{GARSTEC}. Figure~\ref{fig:alpha2.0} shows the computed Kiel diagram with metallicity $\left[\mathrm{Fe/H}\right] = 0.16$ dex, no convective overshooting, and a mixing-length parameter $\alpha_{\mathrm{MLT}}=2.00$. The behaviour of the evolutionary tracks when increasing the mixing-length parameter is similar to decreasing the metallicity (see Fig.~\ref{fig:FeH}) -- the tracks are shifted towards hotter parts of the diagram while the location of the constraint bands within the grid follows the same shift, making the region where the large frequency separation and interferometric radius agree lie within the observed temperatures.
\section{Reconciling parameters}\label{sec:Discussion}
The different trial models of Section~\ref{sec:Results} are used as indicators for the effects on the evolutionary tracks -- and parameters within -- from varying input physics. None of the produced models so far agrees with all parameter constraints at a single point, but combining the various effects can allow for a model in complete agreement. From the trial models it is evident that in order to change the location of the constraint band from the period spacing relative to the other asteroseismic parameters and radius constraints in the Kiel diagram, the inclusion of convective overshooting is a requirement. In this manner a model can be constructed where $\Delta\Pi_1$ and $\Delta\nu$ agrees with the interferometric radius; however the temperature observations are not matched by these models. Since a decrease in metallicity has the same effect on the Kiel diagram as an increase in mixing-length 
-- a shift of the evolutionary tracks towards hotter regions of the diagram -- it is possible to construct a model where the constraints on $\Delta\nu$ and interferometric radius agree with the temperature measurements. A combination of these two effects applied with suitable amounts of overshooting can reconcile discrepancies between the constraints on the physical parameters in the model. 
The primary difference between the effects of decreased metallicity and increased mixing-length, by physically reasonable amounts, is the size of the temperature shift. For a $2\sigma$ decrease in metallicity from the spectroscopic value, as shown in Fig.~\ref{fig:FeH}, the temperature shift is not large enough to make the region where $\Delta\nu$ and radius agree within the $1\sigma$ boundary of the spectroscopic temperature.  
\begin{figure}
\includegraphics[width=\columnwidth]{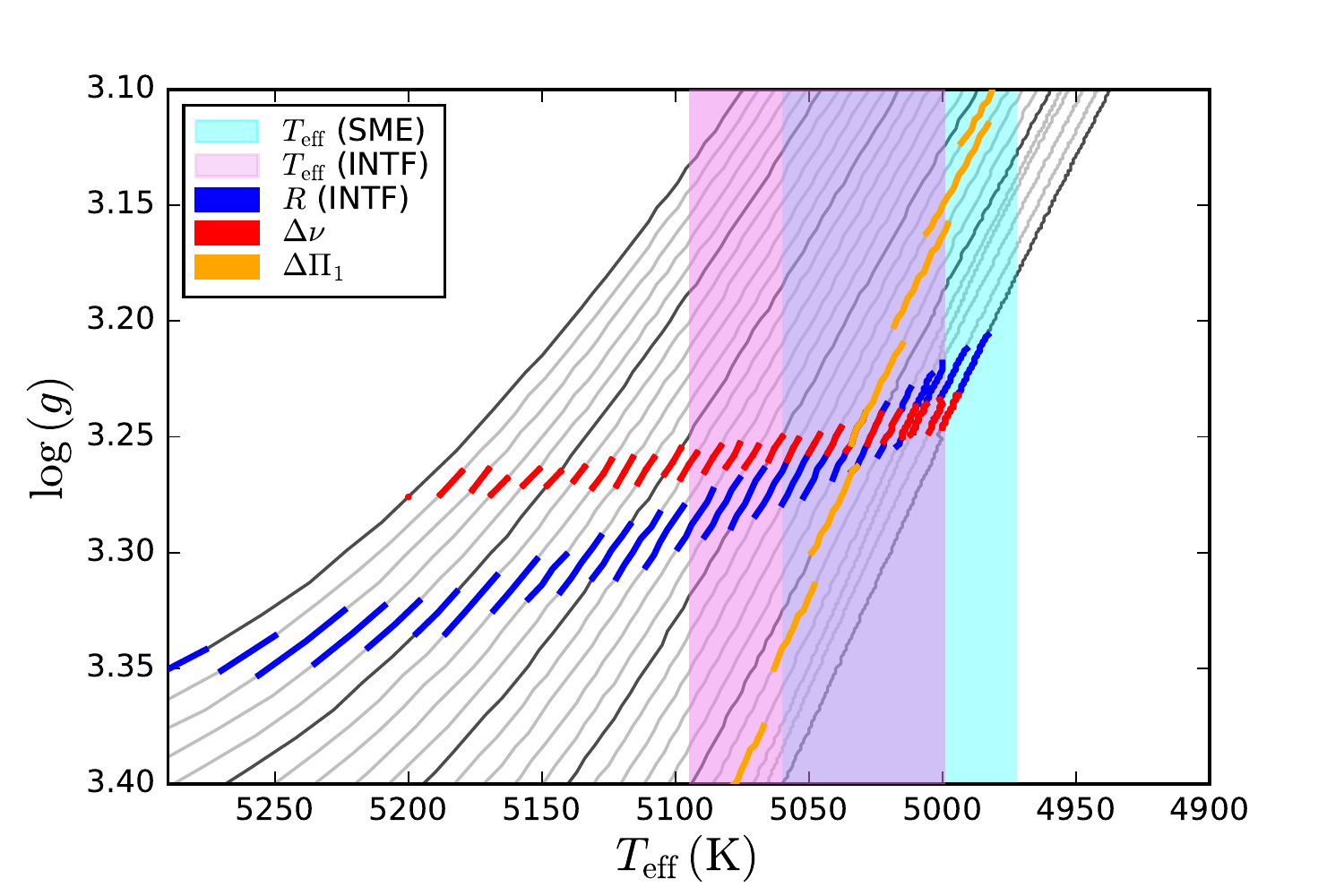}
\caption{Kiel diagram computed with metallicity $\left[\mathrm{Fe/H}\right]=0.16$ dex, convective overshooting included at $f=0.030$, and mixing-length parameter $\alpha_{\mathrm{MLT}}=2.00$. The colour code is as in Fig.~\ref{fig:standard}.}
\label{fig:BestFit_alpha}
\end{figure}

Agreement between all parameter constraints can be obtained by increasing the mixing-length parameter and including convective overshooting. Figure~\ref{fig:BestFit_alpha} shows the Kiel diagram for a model with $\left[\mathrm{Fe/H}\right]=0.16$ dex, $f=0.030$, and $\alpha_{\mathrm{MLT}}=2.00$. For this set of input physics the parameter constraints agree at an evolutionary track of mass $M\simeq 1.60\mathrm{M}_{\sun}$. Thus, within reasonable assumptions on the input of the stellar evolution models, it is possible to reconcile the location of the constraint bands. With this overshoot efficiency and mixing-length parameter we constructed a grid of models spanning a metallicity range between $0.0\leq$[Fe/H]$\leq0.30$, and determined a final set of stellar properties for HD~185351 by reproducing the atmospheric and asteroseismic observables using the BAyesian STellar Algorithm \citep[BASTA,][]{2015MNRAS.452.2127S}. The resulting properties of the star are listed in Table~\ref{tab:parameters}, where we find an excellent agreement between the interferometric radius and the asteroseismically determined one. The small uncertainties in mass and age obtained from the fit are a reflection of the high precision in the \citetalias{Johnson2014} period spacing measurement (at the $0.2\%$ level) as well as the almost perpendicular location of the $\Delta\Pi$ constraint band with respect to the other asteroseismic observables (as depicted in Figs. \ref{fig:standard} and \ref{fig:BestFit_alpha}). Combining these observables effectively breaks the mass degeneracy and allows a much more precise determination of stellar properties. We note however that these uncertainties do not include systematics in the observed asteroseismic values nor the model input physics, and a thorough analysis of these ingredients will be addressed in a subsequent study.
\begin{table}
\centering
\caption{Stellar properties determined using BASTA. As quantities to be reproduced we have considered $\Delta\Pi_1$, $\Delta\nu$, $\nu_\mathrm{max}$, the effective temperature determined from interferometry and the spectroscopic surface abundance. Systematic uncertainties in observed asteroseismic values and model input physics are not included. See text for details.}
\begin{tabular}{cc}
\hline
  & HD~185351 \\
\hline
\smallskip
Mass (M$_\odot$)&  $1.58^{+0.04}_{-0.02}$\\
\smallskip
Radius (R$_\odot$) & $4.92^{+0.15}_{-0.07}$\\
\smallskip
Age (Gyr) & $2.32^{+0.04}_{-0.07}$\\
\smallskip
$\log\,g$ & $3.25^{+0.01}_{-0.02}$\\
\hline
\end{tabular} 
\label{tab:parameters}
\end{table}

This best fit model suggests a mixing-length value larger than the solar-calibrated value. This result is in line with those from 3D simulations of stellar atmospheres where there is a dependence on evolutionary stage and metallicity of the convective efficiency \cite[see][]{Trampedach2014,Magic2015}. The overshooting efficiency parameter $f=0.030$ almost doubles the calibrated value determined by CMD-fitting of M67, where convective overshooting is included to reproduce the hook at the main-sequence turn-off \citep{Magic2010}. The masses of M67 stars in this evolutionary state is $\sim 1.2 \mathrm{M}_{\sun}$, and it is therefore not surprising that a larger efficiency parameter value is found for higher mass stars. Given the exponential nature of the overshoot prescription included in \texttt{GARSTEC} (see Eq.~\ref{eq:overshooting}) the net effect of doubling the overshoot efficiency for a 1.6~M$_\odot$ model is very mild, increasing the maximum convective core size from $m/M=0.09$ to $m/M=0.105$.

Convective core size determination is possible in main-sequence stars using asteroseismology \citep{SilvaAguirre2011,SilvaAguirre2013,Deheuvels2016}, and these results suggest the need of an increasing overshooting efficiency as a function of stellar mass. However solar-like oscillations are detected in very few $1.6\mathrm{M}_{\sun}$ main-sequence stars, making low-luminosity red-giants such as HD~185351 an exciting resource to accurately calibrate the overshooting efficiency.
\section{Conclusions}
In the present work we have shown that variations in the input of the theoretical models can reconcile the discrepancies between interferometric and asteroseismic radius. This has been done using \texttt{GARSTEC} to compute evolutionary tracks with different input physics -- variations in metallicity, alternate values of the mixing-length parameter, and the inclusion of overshooting from convective regions. It has been found that a model with $\left[\mathrm{Fe/H}\right]=0.16$ dex, $f=0.030$, $\alpha_{\mathrm{MLT}}=2.00$, and a stellar mass of $M\simeq 1.60\mathrm{M}_{\sun}$ can reproduce all available observations. Furthermore the inclusion of convective overshooting is found to be crucial in order to match the period spacing constraint with the other parameter constraints in the models.
By analysing stars in a similar evolutionary state as HD 185351 with a wide range in mass it would be possible to calibrate the amount of extra mixing in the core during the MS phase. Such a calibration would allow for an easy and accurate mass estimation from measurements of $\Delta\nu$ and $\Delta\Pi_1$.

We have shown that the systematic errors in the input physics of the adopted models can have a significant effect on the masses and thus need to be considered when estimating systematic errors. The model-dependent input physics, which reconcile all observables by varying the convective core overshooting and mixing-length parameter, suggests a mass of $\simeq 1.60\mathrm{M}_{\sun}$. Thus the conclusion that HD 185351 is a retired A star with a mass in excess of $1.5\mathrm{M}_{\sun}$ still holds.

\section*{Acknowledgements}
Funding for the Stellar Astrophysics Centre is provided by The Danish National Research Foundation (Grant DNRF106). The research is supported by the ASTERISK project (ASTERoseismic Investigations with SONG and Kepler) funded by the European Research Council (Grant agreement no.: 267864). D.H. acknowledges support by the Australian Research Council's Discovery Projects funding scheme (project number DE140101364) and support by the National Aeronautics and Space Administration under Grant NNX14AB92G issued through the Kepler Participating Scientist Program. V.S.A. acknowledges support from VILLUM FONDEN (research grant 10118).




\bibliographystyle{mnras}
\bibliography{ref} 




%
%


\bsp	
\label{lastpage}
\end{document}